\documentclass[%
reprint,
superscriptaddress,
groupedaddress,
amsmath,amssymb,
aps,
prl,
]{revtex4-2}

\usepackage[table]{xcolor}
\usepackage{xcolor}
\usepackage{amsmath}
\usepackage{graphicx}
\usepackage{dcolumn}
\usepackage{bm}
\usepackage{hyperref}

\newcommand{\kvec}{\mathbf{k}}

\newcommand{\qvec}{\mathbf{q}}
\newcommand{\aF} {\alpha^{2}F(\omega)}
\begin{document}

\title{First-Principles Calculation of Superconducting $T_c$ in Superhard B-C-N Metals}

\author{Adam D. Smith}
\email{smitha20@uab.edu}
\affiliation{Department of Physics, University of Alabama at Birmingham, Birmingham, Alabama 35294, USA}

\author{Yogesh K. Vohra}
\affiliation{Department of Physics, University of Alabama at Birmingham, Birmingham, Alabama 35294, USA}

\author{Cheng-Chien Chen}
\email{chencc@uab.edu}
\affiliation{Department of Physics, University of Alabama at Birmingham, Birmingham, Alabama 35294, USA}

\date{\today}%

\begin{abstract}
We perform first-principles electron-phonon calculations to evaluate the superconducting transition temperature $T_c$ for ternary superhard metals B$_2$C$_3$N and B$_4$C$_5$N$_3$. An ambient-pressure $T_c$ of $\sim 40$ K and $\sim 20$ K is obtained respectively for B$_2$C$_3$N and B$_4$C$_5$N$_3$ from the Eliashberg gap equations. The relatively high $T_c$ of these compounds is due in part to their high Debye temperatures associated with superhardness. The materials under study are potentially synthesizable, as their formation energies are comparable to those of other recently synthesized superhard B-C-N compounds. Therefore, studying superhard metals could hold the promise of realizing new higher-$T_c$ superconductors at ambient pressure.
\end{abstract}


\maketitle

\textit{Introduction --}
A recent resurgence of interests in studying phonon-mediated Bardeen-Coopper-Schrieffer (BCS) superconductors has largely resulted from the successful prediction and realization of superconducting high-pressure hydrides H$_3$S and LaH$_{10}$~\cite{li2014metallization, drozdov2015conventional,peng2017hydrogen, drozdov2019superconductivity,somayazulu2019evidence}.
These achievements have greatly enhanced the prospects of room-temperature superconductivity. However, the requirement for high external pressures in the megabar (Mbar or 100 GPa) range to synthesize and maintain superconducting states in these materials limits their practical applications. In addition to hydrogen-rich materials~\cite{zurek2019high, chen2021high, dolui2024feasible}, other high-pressure systems based on light elements beryllium (Be), boron (B), carbon (C), and/or nitrogen (N) also have been studied~\cite{li2019hard, hutcheon2020predicting,gao2021phonon,lim2021high,hai2023superconductivity, sanna2024prediction,tomassetti2024prospect}. Searching for new, synthesizable superconductors with higher $T_c$ at reduced pressure would promise to revolutionize the energy, transportation, and information technologies~\cite{de2021materials, mitchell2021superconductors, yazdani2023roadmap}.

Generally, the $T_c$ of phonon-mediated (BCS) superconductors increases in materials with a large electron-phonon coupling strength $\lambda$, and/or a high Debye temperature $\Theta_D$~\cite{mcmillan1968transition,allen1975transition}.
Recent theoretical studies have suggested that $T_c \le A\Theta_D$ for BCS superconductors, with the pre-factor $A \sim 0.1$~\cite{esterlis2018bound, smith2023machine}. $\Theta_D$ is typically elevated in highly compressed materials due to pressure-induced hardening. In the absence of external pressure, high $\Theta_D$ can occur in materials with superior mechanical properties, such as those exhibiting superhardness.
In Vickers indentation test~\cite{smith1922accurate}, hardness $H_V$ is determined by the ratio of the applied force to the resulting indentation area. A material is called superhard when its $H_V \ge 40$ GPa~\cite{yeung2016ultraincompressible}. At ambient pressure, diamond is the known hardest material with $H_V \sim 100$ GPa~\cite{haines2001synthesis} and a high $\Theta_D \sim 2200$ K~\cite{gschneidner1964physical}. However, diamond fails to become a BCS superconductor due to its large insulating gap.
On the other hand, a metallic superhard compound, BC$_5$, was theoretically suggested to show a relatively high $T_c$ in the range $\sim 11 - 45$ K, depending on the underlying crystal structures~\cite{BC5Superconductor_Calandra2008,yao2009crystal,li2010superhard}. BC$_5$ with a local covalently-bonded structure similar to cubic diamond can exhibit a high hardness $H_V \ge 60$ GPa~\cite{solozhenko2009ultimate, baker2018computational}. The $T_c$ for other binary superhard compounds like BC$_3$ and BC$_7$ were also computed to be between $\sim 11 - 23$ K~\cite{liu2011superhard,xu2011prediction}.
It remains an important experimental task to verify if these materials can be synthesized in the proposed structures, and if they can superconduct at the high theoretical $T_c$.

\begin{figure}[!t]
\centering
\includegraphics[width=0.5\textwidth]{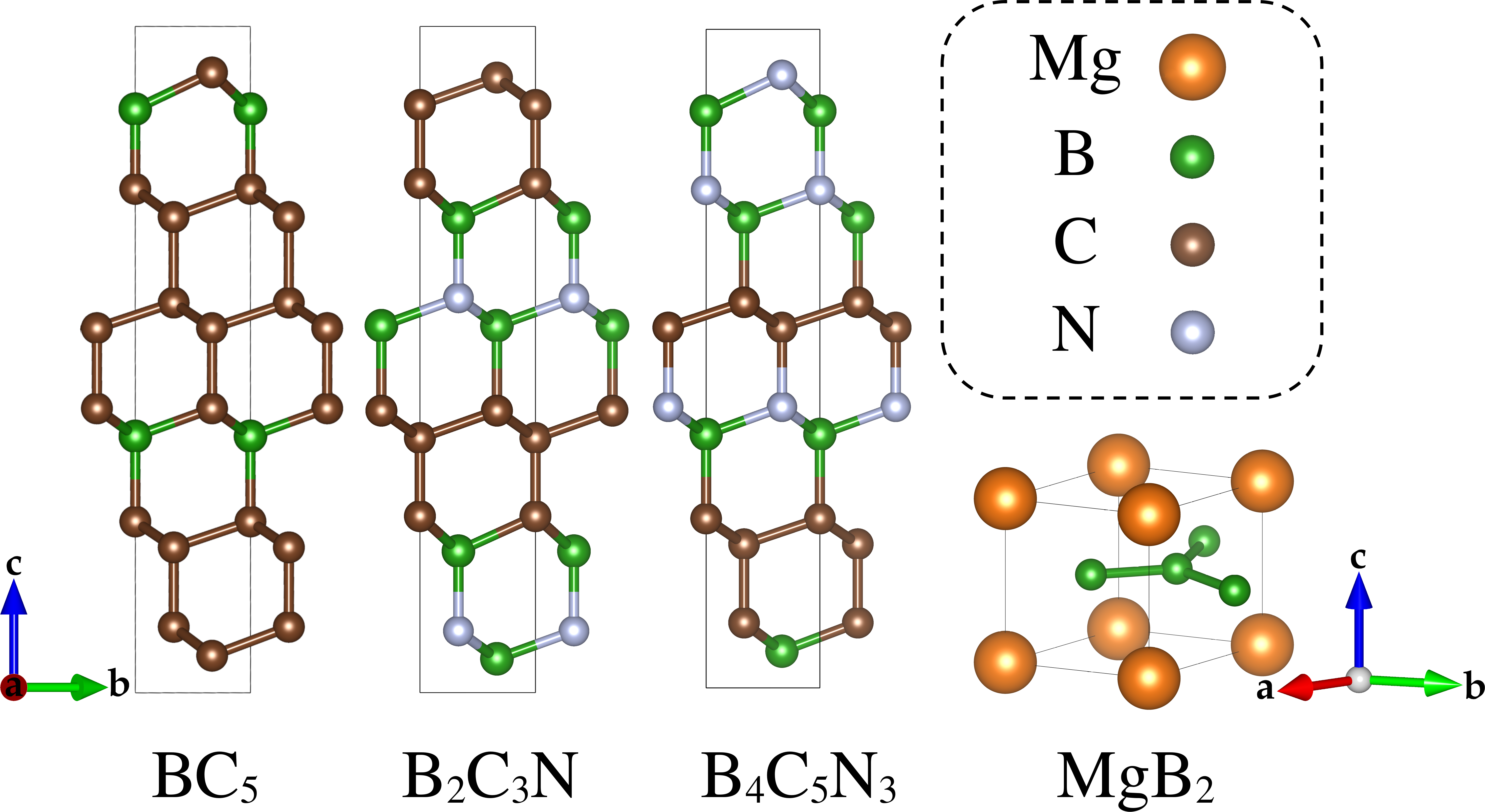}
\caption{Schematic crystal structures under study. BC$_{5}$, B$_{2}$C$_{3}$N, and B$_{4}$C$_{5}$N$_{3}$ with a 12-atom unit cell belong to trigonal symmetry of space group P3m1 (No. 156). MgB$_{2}$ belongs to hexagonal symmetry of space group P6/mmm (No. 191).
The crystal axes shown on the left correspond to the B-C and B-C-N systems, while those on the right correspond to MgB$_2$.
}
\label{fig:systems}
\end{figure}

In general, adding an additional element provides more degrees of freedom in the phase space for exploring new materials. In this work, we perform first-principles electron-phonon calculations to study the superconducting $T_c$ of B$_2$C$_3$N and B$_4$C$_5$N$_3$. Stable superhard ($H_V \ge 40$ GPa) metallic structures with these ternary chemical compositions (as shown in Fig. \ref{fig:systems}) were recently predicted by machine-learning and evolutionary structure searches~\cite{chen2021machine}.
These materials exhibit multiple nested Fermi surface sheets, which may exhibit anisotropic electron-phonon couplings and necessitates solutions of the full Eliashberg equations for a precise determination of $T_c $. The  anisotropic calculations indicate $T_c$ values of $\sim 40$ K and $\sim 20$ K for B$_2$C$_3$N and B$_4$C$_5$N$_3$, respectively.
Notably, the theoretical formation energy of B$_2$C$_3$N and B$_4$C$_5$N$_3$ is comparable to that of previously synthesized ternary superhard materials such as BC$_2$N and BC$_{10}$N~\cite{solozhenko2001synthesis, chakrabarty2024novel}.
Because the theoretical formation energies of all these B-C-N materials lie within 0.2 eV/atom~\cite{chen2021machine}, it is feasible that the B-C-N materials under study could be synthesized~\cite{aykol2018thermodynamic}, using high-pressure high-temperature (HPHT) techniques or microwave plasma chemical vapor deposition (MPCVD).
Consequently, superhard metals could offer an intriguing new avenue for achieving higher-$T_c$ BCS superconductors under ambient pressure.

\textit{Methods --} First-principles density functional theory (DFT) calculations are based on the Quantum ESPRESSO software (version 7.2)~\cite{QuantumEspressoPaper_Gianozzi2017, QuantumEspressoPaper2_Gianozzi2009}, and electron-phonon calculations are performed with the EPW code (version 5.7)~\cite{EPWpaper_Lee2023,EPW_Giustino2017,Elphwann_Giustino2007}.
In addition to B$_2$C$_3$N and B$_4$C$_5$N$_3$, we perform benchmark calculations for BC$_5$ and MgB$_2$ (the current $T_c$ record-holder among ambient-pressure BCS superconductors).
Figure \ref{fig:systems} shows the structures under study.
The BC$_{5}$, B$_2$C$_3$N, and B$_4$C$_5$N$_3$ structures are sourced from their respective papers~\cite{chen2021machine, BC5Superconductor_Calandra2008} and arranged into 12-atom unit cells. The structure for MgB$_{2}$ is sourced from the Materials Project~\cite{MaterialsProject_Jain2013}. 
The Monkhorst-Pack sampling scheme~\cite{MonkhorstPack1976} is used with a $\Gamma$-centered $\mathbf{k}$-point mesh of $28\times 28\times 5$ for the B-C and B-C-N materials, and a mesh of $21\times 21\times 21$ for MgB$_2$.
A Marzari-Vanderbilt smearing~\cite{MarzariSmearing1999} of 6 mRy is employed. The convergence criteria of self-consistent and structural relaxation calculations are set to $10^{-8}$ Ry per unit cell and $10^{-6}$ Ry per Bohr radius, respectively. We adopt a wavefunction kinetic energy cutoff of 100 Ry, which suffices to converge the DFT total energy difference within $10^{-6}$ Ry per atom. For each crystal structure, we fully relax the lattice parameters and atomic positions without pressure. Afterwards, we compute the electronic, phonon, and electron-phonon properties.
All calculations use norm-conserving pseudopotentials from the standardized \textit{PseudoDojo} library~\cite{Pseudodojo_VanSetten2018} and the Perdew-Burke-Ernzerhof (PBE) exchange-correlation functional~\cite{PBE_Perdew1996}, with the following valence states for each element: B:$2s^22p^1$, C:$2s^22p^2$, N:$2s^22p^3$, and Mg:$2s^22p^63s^2$.
The dynamical matrices and linear variation of the self-consistent potential are calculated with density-functional perturbation theory (DFPT)~\cite{DFPT_Baroni2001}.
Here, the same $\mathbf{k}$-point sampling meshes described above are used. The $\mathbf{q}$-point sampling mesh is $9\times 9\times 3$ for the B-C and B-C-N materials, and $7\times 7\times 7$ for MgB$_2$. These meshes are sufficiently converged for the systems under study.

The EPW calculations use Wannier interpolation on uniform $\Gamma$-centered $\mathbf{k}$-grids with the Wannier90 code~\cite{WannierEcosystem_Marrazzo2023,Wannier90_Pizzi2020}. In the cases of BC$_{5}$, B$_{2}$C$_{3}$N, and B$_{4}$C$_{5}$N$_{3}$, we consider $2p$ orbitals for every B and C atom as initial projections for the maximally localized Wannier functions~\cite{MLWF_Marzari2012}; in MgB$_2$, Mg $3s$ and B $2p_z$ orbitals are considered.
After Wannierization, we perform electron-phonon computation on dense $70\times 70 \times 14$ ($70\times 70\times 70$) $\kvec$ and $35 \times 35\ \times 7$ ($35 \times 35 \times 35$) Fourier-interpolated $\qvec$ grids for the B-C and B-C-N systems (MgB$_2$) to evaluate the Eliashberg spectral function $\aF$ and the isotropic electron-phonon coupling $\lambda$.
The anisotropic Eliashberg equations are solved on the same dense interpolated $\kvec$ and $\qvec$ grids with imaginary Matsubara frequencies~\cite{AnisotropicEliashberg_Margine2013}. The solutions are brought to the real axis with Padé approximants. The crystal structures and Fermi surfaces are visualized respectively by VESTA~\cite{VESTA_Momma2011} and FermiSurfer~\cite{FermiSurfer_Kawamura2019} softwares.

\begin{figure}[!t]
\centering
\includegraphics[width=0.5\textwidth]{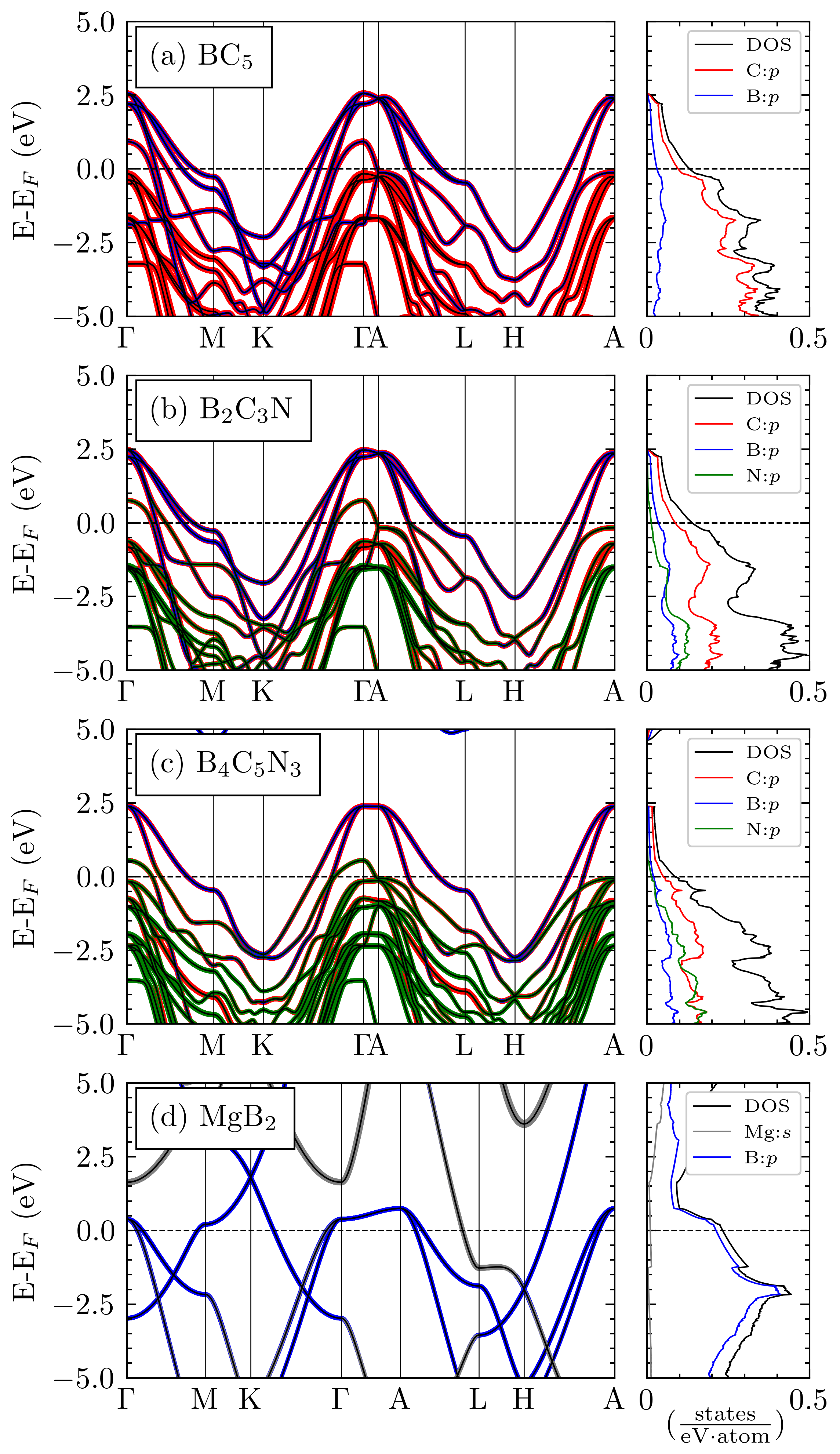}
\caption{Electronic band structures  along high-symmetry paths [left] and density of states [right] for (a) BC\(_5\), (b) B\(_2\)C\(_3\)N, (c) B\(_4\)C\(_5\)N\(_3\), and (d) MgB\(_2\). The energy is plotted relative to the Fermi level (\(E_F\)) indicated by the horizontal dashed line. The B-C and B-C-N compounds exhibit band structures similar to diamond but are effectively hole doped.}
\label{fig:bands}
\end{figure}

\begin{figure*}[!t]
\centering
\includegraphics[width=0.85\textwidth]{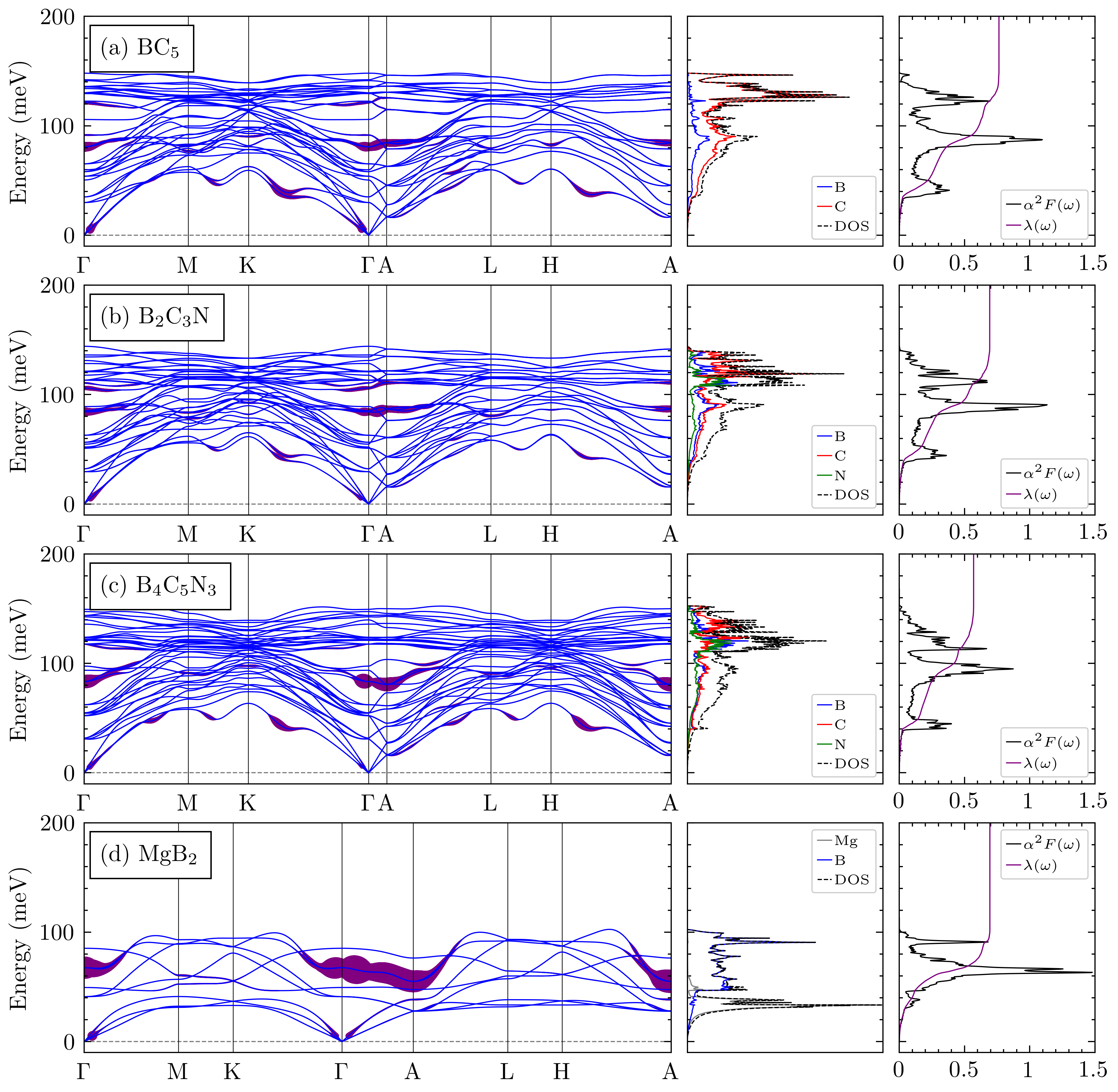}
\caption{Phonon dispersions along high-symmetry paths [blue curves, left panels] for (a) BC$_{5}$, (b) B$_{2}$C$_{3}$N, (c) B$_{4}$C$_{5}$N$_{3}$, and (d) MgB$_{2}$.
A Gaussian-filtered average of mode-resolved electron-phonon coupling $\lambda_{\qvec\nu}$ is plotted in purple, with the width displaying its intensity. The middle panels show the atom-projected phonon density of states (PHDOS). The right panels show the Eliashberg spectral function $\aF$ and the isotropic electron-phonon coupling $\lambda$.
}
\label{fig:phonons_a2f}
\end{figure*}

\textit{Results and Discussion --}
Figure \ref{fig:bands} shows the atom-projected electronic band structures and density of states (DOS).
The DOS spectra are obtained directly from DFT calculations using the tetrahedron integration method. For consistency with the EPW calculations, the DFT DOS are evaluated on the same dense $\mathbf{k}$-point meshes used for fine-grid interpolation of the electronic structure: $70\times 70\times 14$ for the B-C and B-C-N materials, and $70\times 70\times 70$ for MgB$_2$.
BC$_{5}$, B$_{2}$C$_{3}$N, and B$_{4}$C$_{5}$N$_{3}$ all exhibit local covalently bonded atomic arrangements similar to diamond. Therefore, they have similar band structures [Figs. \ref{fig:bands}(a)-(c)], but are effectively hole doped with the Fermi level ($E_F$) residing within the valence bands.
In BC$_5$ and B$_2$C$_3$N, the atomic weights near $E_F$ of C and B atoms dominate over that from N. In B$_4$C$_5$N$_3$, while C still dominates, the contributions from B and N are comparable.
Specifically, BC$_{5}$ and B$_{2}$C$_{3}$N feature five bands crossing the $E_F$ with comparable electron DOS at the Fermi level ($N_F$). In contrast, B$_{4}$C$_{5}$N$_{3}$ shows only three bands crossing the $E_F$ with a much lower $N_{F}$. This variation in $N_F$ between B$_{2}$C$_{3}$N and B$_{4}$C$_{5}$N$_{3}$ is expected, due to the differing hole doping levels based on the valence electron counts of B and N atoms. A larger difference in the B-to-N ratio will result in stronger hole doping. A higher $N_F$ can lead to stronger electron-phonon coupling. Therefore, although the theoretical Debye temperatures of B$_{2}$C$_{3}$N and B$_{4}$C$_{5}$N$_{3}$ are comparable~\cite{chen2021machine}, B$_{2}$C$_{3}$N can exhibit a higher $T_c$, as discussed later.

The electronic structure for MgB$_2$ is shown in Fig. \ref{fig:bands}(d). There is a notable difference in the band topology between MgB$_{2}$ and the other compounds under study. MgB$_{2}$ is known to exhibit a complex mixture of $\sigma$ and $\pi$ orbitals, which manifest as both 2D and 3D Fermi surface sheets~\cite{choiAnisotropicEliashbergTheory2006}.
Conversely, $p$ orbitals dominate the bands near the $E_F$ in BC$_5$, B$_{2}$C$_{3}$N, and B$_{4}$C$_{5}$N$_{3}$.
The variation in $N_F$ suggests differences in how electrons interact with phonons in these materials, which play a vital role in determining their phonon-mediated superconductivity.

Figure \ref{fig:phonons_a2f} left panels show phonon dispersions along high-symmetry paths of the materials under study.
The absence of negative phonon modes confirms the dynamic stability of all the materials at ambient pressure. As seen in Figs. \ref{fig:phonons_a2f}(a)-(c), the phonon spectra of BC$_{5}$, B$_{2}$C$_{3}$N, and B$_{4}$C$_{5}$N$_{3}$ exhibit a prominent contribution from high-frequency optical phonons, which is a common feature of covalently bonded superhard materials like diamond.
In contrast, MgB$_2$ displays a smaller phonon bandwidth and a quite different band topology, as seen in Fig. \ref{fig:phonons_a2f}(d).
There is also a major difference in the distribution of the mode-resolved electron-phonon coupling $\lambda_{\qvec\nu}$~\cite{EPWpaper_Lee2023,EPW_Giustino2017,Elphwann_Giustino2007}:
\begin{equation}
\lambda_{\qvec\nu} = \frac{\hbar}{2 N_{F}} \sum_{mn, \kvec} W_{\kvec} \frac{1}{\omega_{\qvec \nu}^{2}}|g_{mn,\nu}|^{2} \delta(\epsilon_{n\kvec}) \delta(\epsilon_{m\kvec+\qvec}),
\end{equation}
where $W_{\kvec}$ is a weight factor equivalent to the reciprocal number of $\kvec$ points on the Fermi surface; $\omega_{\qvec \nu}$ is the phonon energy of mode $\nu$ at phonon wavevector $\qvec$; $g_{mn, \nu}$ is the electron-phonon coupling matrix element between electron bands $m$ and $n$ with phonon mode index $\nu$; $\epsilon_{n\kvec}$ and $\epsilon_{m\kvec+\qvec}$ are the electron eigen-energies with band indices $n$ and $m$ at momenta $\kvec$ and $\kvec + \qvec$, respectively. 

\begin{figure*}[!t]
\centering
\includegraphics[width=0.75\textwidth]{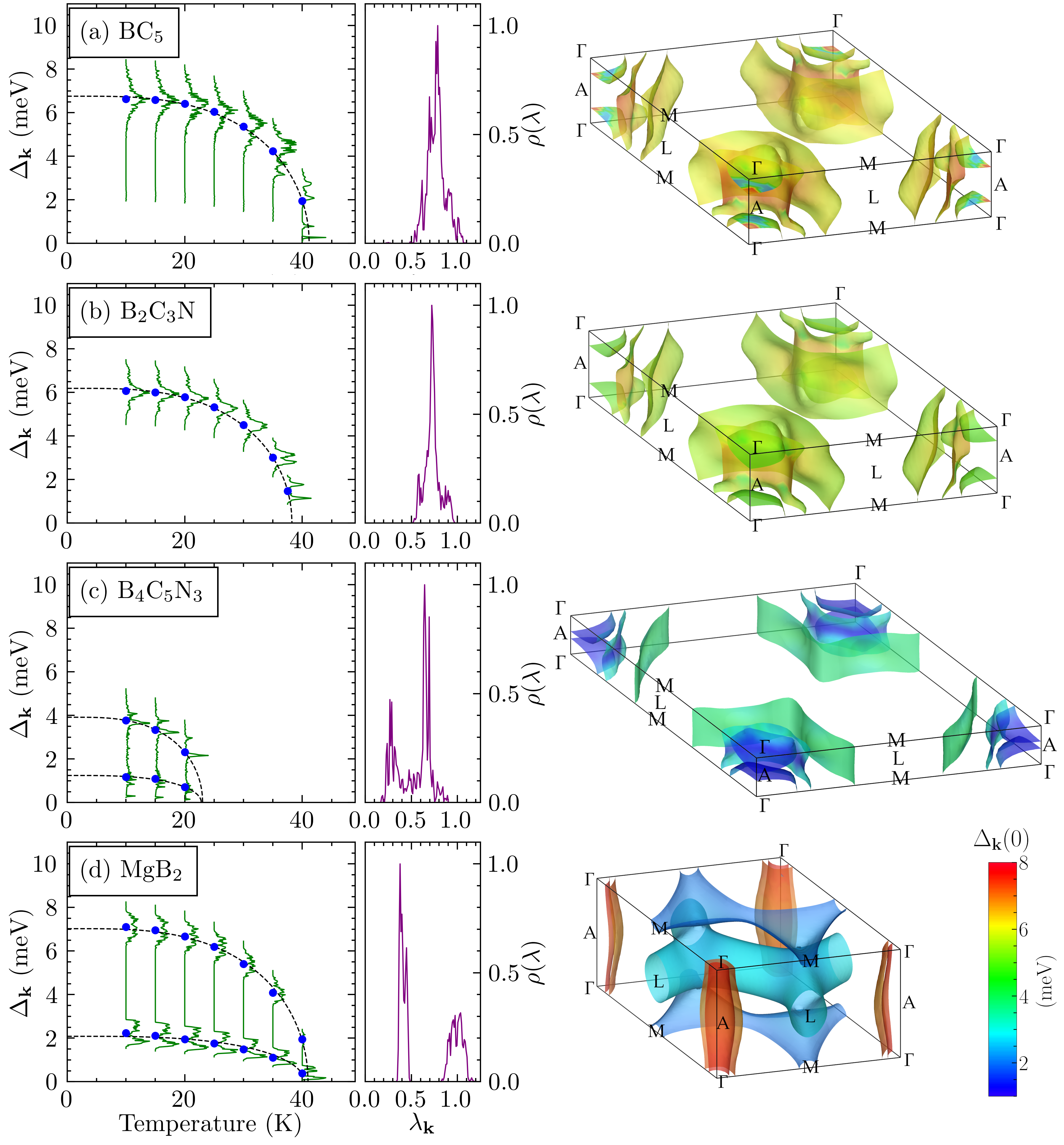}
\caption{Superconducting gap $\Delta_{\mathbf{k}}$ as a function of temperature [green curves, left panels] computed by anisotropic Migdal-Eliashberg theory for (a) BC$_{5}$, (b) B$_{2}$C$_{3}$N, (c) B$_{4}$C$_{5}$N$_{3}$, and (d) MgB$_{2}$. The blue dots are first-moment averages of $\Delta_{\mathbf{k}}$, and they are fitted to the BCS gap equation (black dashed line) to determine the $T_c$. The middle panels show the distribution $\rho(\lambda)$ of the momentum-resolved electron-phonon coupling on the Fermi surface $\lambda_{\kvec}$.
The false-color intensity plots on the right represent $\Delta_{\mathbf{k}}$ values projected onto the Fermi surface.}
\label{fig:gaps_fs}
\end{figure*}

As seen in the purple shaded areas of Figs. \ref{fig:phonons_a2f}(a)-(c), $\lambda_{\qvec\nu}$ in the B-C and B-C-N systems predominantly arises from the higher-energy optical phonons around 80 meV along the $\Gamma-A$ direction of the Brillouin zone. The atom-projected phonon DOS (Fig. \ref{fig:phonons_a2f} middle panels) indicate that the modes contributing to the strong electron-phonon interactions are mainly associated with B and C atoms. A closer examination of the phonon eigenstates reveals that these modes involve predominantly in-plane ($x-y$) bond-stretching motions between B and C atoms in opposite directions.
There are additional contributions to $\lambda_{\qvec\nu}$ from the lower-energy acoustic phonons near 50 meV. These low-energy contributions are associated with notable phonon softenings and enhanced $\lambda_{\qvec\nu}$ along the $K-\Gamma$ and $H-A$ Brillouin zone paths. As seen later in Fig. \ref{fig:gaps_fs} insets, the Fermi surface topologies of the materials under study provide good nesting conditions, which can induce phonon softenings and Kohn anomalies to facilitate a stronger electron-phonon coupling~\cite{aynajian2008energy,zhang2019pressure, yang2022kohn} 
As a comparison, $\lambda_{\qvec\nu}$ in MgB$_2$ is dominated by the phonon modes near $\sim 60-80$ meV [Fig. \ref{fig:phonons_a2f}(d)]. Corresponding optical-phonon softenings and Kohn anomalies also have been discussed in the context of MgB$_2$~\cite{alarco2015phonon,mackinnon2017phonon,johansson2022effect}.

The mode-resolved electron-phonon coupling $\lambda_{\qvec\nu}$ (already averaged over $\kvec$) can be further averaged over $\qvec$ to obtain the isotropic electron-phonon coupling $\lambda$:
\begin{equation}
    \lambda = \frac{1}{N_{\qvec}} \sum_{\qvec \nu} \lambda_{\qvec \nu} = 2 \int \frac{\aF}{\omega} d \omega.
\end{equation}
Here, $N_{\qvec}$ is the total number of points in the fine $\qvec$ grid. $\aF$ is the Eliashberg spectral function defined as:
\begin{equation}
    \aF = \frac{1}{2 N_{\qvec}} \sum_{\qvec \nu} \omega_{\qvec \nu} \lambda_{\qvec \nu} \delta(\omega - \omega_{\qvec \nu}).
\end{equation}
The atom-projected phonon density of states (PHDOS), $\aF$, and $\lambda$ are also plotted in Fig. \ref{fig:phonons_a2f}.
Compared to MgB$_{2}$, BC$_{5}$ and B$_{2}$C$_{3}$N achieve a comparable $\lambda \sim 0.7$ due to contributions from high-energy phonon modes, which compensate for their lower electron DOS at the $E_F$ ($N_F$).
The lattice stiffness and superhardness of the B-C and B-C-N compounds play crucial roles in sustaining strong electron-phonon coupling at these higher phonon energies. On the other hand, B$_4$C$_5$N$_3$ exhibits a slightly smaller $\lambda \sim 0.6$, likely due to an even smaller $N_F$.

Often times, it suffices to estimate the $T_c$ of phonon-mediated superconductors by Allen-Dynes equation~\cite{allen1975transition}:
\begin{equation} \label{eqn:ad_tc}
    T_{c} = \frac{\omega_{log}}{1.2} \mathrm{exp} \left[ - \frac{1.04(1+\lambda)}{\lambda-\mu^{*} (1+0.62\lambda)}\right].
\end{equation}
The logarithmic average frequency $\omega_{log}$ is defined as:
\begin{equation} \label{eqn:omega_log}
    \omega_{log} = \mathrm{exp} \left[ \frac{2}{\lambda} \int_{0}^{\infty} d\omega 
    \frac{\aF}{\omega} \log(\omega)\right].
\end{equation}
The Coulomb pseudopotential $\mu^{*}$ entering the Allen-Dynes equation is an effective parameter describing the effects of Coulomb repulsion on superconductivity. 
In the literature~\cite{EPW_Giustino2017}, a relatively wide range of $\mu^{*} \sim 0.1-0.3$ is commonly considered when evaluating the superconducting $T_c$. As shown below, our choice of $\mu^{*} = 0.1$ yields a theoretical $T_c \sim40$ K for MgB$_2$, which is in close agreement with experiment. However, because $\mu^{*}$ can vary between compounds, the $T_c$ computed with $\mu^{*} = 0.1$ may be overestimated and thus likely represents as an upper bound for the actual experimental $T_c$ of B-C-N materials.

\begin{table}[h]
\setlength{\tabcolsep}{5pt} 
\begin{center}
\begin{tabular}{c | c c c c c c}
\hline
\hline
System & $N_{F}$ & $\lambda$  & $\omega_{log}$ & $T_{c,AD}$ &  $T_{c,iME}$ & $T_{c,aME}$  \\ 
\hline
BC$_{5}$               & 0.133 & 0.764 & 70.09 & 34.6 & 41.5 & 41.2 \\ \hline  
B$_{2}$C$_{3}$N        & 0.146 & 0.693 & 74.32 & 29.3 & 37.3 & 38.2 \\ \hline  
B$_{4}$C$_{5}$N$_{3}$  & 0.081 & 0.570 & 74.98 & 16.9 & 17.9 & 22.9 \\ \hline 
MgB$_{2}$              & 0.232 & 0.694 & 56.97 & 22.5 & 22.9 & 40.7 \\ \hline 
\hline
\end{tabular}
\end{center}
\caption{First-principles results for the electron density of states at the Fermi level $N_{F}$ [in units of states/(eV$\cdot$atom)], the dimensionless isotropic electron-phonon coupling $\lambda$, the logarithmic average frequency $\omega_{log}$ (in meV), and the superconducting transition temperature computed by the Allen-Dynes equation $T_{c,AD}$ (in Kelvin).
The superconducting transition temperatures $T_{c,iME}$ (in Kelvin) and $T_{c,aME}$ (in Kelvin) are obtained by fitting the BCS gap equation to solutions of the isotropic and anisotropic Migdal-Eliashberg theory, respectively. All $T_c$ values are calculated with $\mu^{*}=0.1$.
}
\label{tab:SC_properties}
\end{table}

Table $\ref{tab:SC_properties}$ summarizes the calculated $\lambda$, $\omega_{log}$, and the superconducting transition temperature $T_{c, AD}$ using the Allen-Dynes equation (Eq. \ref{eqn:ad_tc}) with $\mu^*=0.1$. BC$_{5}$ achieves a much higher $T_{c,AD}$ compared to the other compounds under study.
However, the isotropic approximation of $\lambda$ does not account for the anisotropy in electron-phonon interactions. 
In general, the presence of multiple nested Fermi surface sheets may sustain anisotropic superconducting gaps. Additionally, the computed $T_{c,AD}$ for MgB$_2$ is only 22.5 K, much lower than the experimental value of $\sim 40$ K. This discrepancy suggests that solutions of the full Eliashberg gap equations are required.

The Migdal-Eliashberg (ME) theory using Wannier functions~\cite{AnisotropicEliashberg_Margine2013} takes into account momentum-dependent interactions on the Fermi surface and variations in electron-phonon coupling across different phonon modes. By solving both the isotropic and anisotropic ME equations on fine momentum and frequency grids, we can determine the superconducting gaps on the Fermi surface to more accurately evaluate $T_c$.
Figure \ref{fig:gaps_fs} shows the resulting anisotropic gap functions and their projections onto the Fermi surfaces.
BC$_5$ and B$_2$C$_3$N exhibit a similar Fermi surface topology and a single superconducting gap [Figs. \ref{fig:gaps_fs}(a)-(b)].
In contrast, B$_4$C$_5$N$_3$ manifests two gaps similar to MgB$_2$ [Figs. \ref{fig:gaps_fs}(c)-(d)].
B$_{4}$C$_{5}$N$_{3}$ has a Fermi surface structure similar to BC$_{5}$ and B$_{2}$C$_{3}$N, but with a flatter and more pronounced hexagonal central band and cylindrical outer band.
In MgB$_{2}$, the two gaps originate from the distinct 2D and 3D Fermi surfaces. Its $\sigma$ bands are quasi-two-dimensional and show stronger electron-phonon coupling with low-energy phonon modes, causing a higher superconducting gap. The $\pi$ bands are more three-dimensional, showing a weaker coupling and a lower superconducting gap. 

The superconducting $T_c$ of the ME theory can be obtained by fitting the leading edge of the superconducting gap on the Fermi surface as a function of temperature, $\Delta_{\kvec}(T)$, to the analytical BCS gap equation:
\begin{equation}
\Delta_{\kvec}(T) = \Delta_{\kvec}(T=0) \tanh\left(1.74 \sqrt{\frac{T_{c}}{T} - 1}\right).
\end{equation}
The fitted superconducting transition temperatures $T_{c,iME}$ and $T_{c,aME}$ respectively for the isotropic and anisotropic ME theory are summarized in Table \ref{tab:SC_properties}. $T_{c,iME}$ and $T_{c,aME}\sim 40 K$ are close in BC$_5$ and B$_2$C$_3$N, indicating that these two materials can be characterized as isotropic superconductors. However, $T_{c,iME}$ and $T_{c,aME}$ in B$_4$C$_5$N$_3$ and MgB$_2$ differ more substantially, suggesting strong anisotropy~\cite{choi2002first,mazin2004comment,PhysRevB.69.056502,choi2006anisotropic,ummarino2005two,de2010electron}. In particular, $T_{c,aME}$ for MgB$_2$ is 40.7 K, which agrees well with the experiments. Our results indicate the importance of investigating the full Eliashberg gap equations for accurate $T_c$ determination.

Finally, the anisotropic nature also can be seen from the distribution $\rho(\lambda)$ of the momentum-resolved electron-phonon coupling on the Fermi surface $\lambda_{\kvec}$, as shown in the middle panels of Fig. \ref{fig:gaps_fs}.
In BC$_5$ and B$_2$C$_3$N [Figs. \ref{fig:gaps_fs}(a)-(b)], $\lambda_{\kvec}$ is peaked around 0.7 and 0.8. In B$_4$C$_5$N$_3$ and MgB$_2$ [Figs. \ref{fig:gaps_fs}(c)-(d)], however, $\lambda_{\kvec}$ shows clearly a two-peak structure. From the false-color intensity plots for the Fermi-surface projected superconducting gaps [Fig. \ref{fig:gaps_fs} right panels], BC$_5$ and B$_2$C$_3$N show a more uniform color. In contrast, B$_4$C$_5$N$_3$ and MgB$_2$ exhibit two more distinct colors associated with anisotropic gaps. Such an anisotropy can lead to a highly underestimated $T_c$ in the isotropic approximation of the electron-phonon coupling.

\textit{Conclusion --}
We have performed detailed first-principles electron-phonon calculations to study superhard metals BC$_{5}$, B$_{2}$C$_{3}$N, and B$_{4}$C$_{5}$N$_{3}$. Their superconducting transition temperatures computed using Migdal-Eliashberg theory were found comparable to that of MgB$_2$, which, two decades after its discovery, remains the $T_c$ record holder among ambient-pressure BCS superconductors. The B-C-N materials under study have relatively low formation energies and can potentially be made by high-pressure high-temperature (HPHT) synthesis or microwave plasma chemical vapor deposition (MPCVD). Further theoretical explorations of $T_c$ in different superhard metals via the Eliashberg gap equations, alongside experimental investigations of their synthesizability and superconducting properties, are important future research areas for discovering new higher-$T_c$ phonon-mediated superconductors.

\section*{acknowledgments}
The research is supported by the U.S. National Science Foundation (NSF) Award Nos. DMR-2310526 and OIA-2148653. C.-C.C. also acknowledges support from NSF Award No. DMR-2142801. The calculations utilized the Frontera computing system at the Texas Advanced Computing Center made possible by NSF Award No. OAC-1818253.


\bibliography{apssamp}

\end{document}